\documentclass[12pt]{iopart}
\usepackage{iopams} 
\usepackage{graphicx}

\begin{document}

\title[Three-dimensional diffusion]{Three dimensional diffusion with helical persistence}

\author{Hern\'an Larralde$^1$
and Fran\c cois Leyvraz$^1$ \footnote{also at Centro Internacional de Ciencias, Cuernavaca, Morelos M\'exico}}

\address{$^1$ Instituto de Ciencias F\'\i sicas---Universidad Nacional Aut\'onoma de M\'exico, 
Cuernavaca, Morelos, M\'exico}
\eads{\mailto{hernan@fis.unam.mx},\mailto{leyvraz@fis.unam.mx}}

\begin{abstract}
We formulate the the problem of persistent diffusion in three
dimensions from the perspective of the Frenet--Serret equations. In
contrast to one and two dimensional systems, in three dimensions
persistent diffusion is, in general, a third order process. In this
paper we derive a Fokker--Planck equation for the process and we
calculate its effective diffusion constant. We also provide
expressions for the asymptotic average displacement of the walk, as
well as explicit expressions for the Fourier--Laplace transform of the
correlations between the tangent, normal and binormal vectors of the
motion. 

\end{abstract}

%Uncomment for PACS numbers title message
%\pacs{00.00, 20.00, 42.10}
% Keywords required only for MST, PB, PMB, PM, JOA, JOB? 
%\vspace{2pc}
\noindent{\it Keywords}: diffusion, non-markovian processes, helical motion, swimming
% Uncomment for Submitted to journal title message
%\submitto{\JPA}
% Comment out if separate title page not required
\maketitle

\section{Introduction}
Persistent random walks have been proposed as models for a wealth of
physical and biological systems \cite{CrenI, CrenII, CrenIII, Flory,
  Julicher, Becker} . In the simplest scenario, this
process is characterized by the tendency to continue moving in the
direction of the last step. In certain circumstances, this tendency
can be thought of a kind of inertial effect that keeps the particle
moving in the direction of motion \cite{Weiss}. On the other hand, in the context
of biology, the tendency may arise from the fact that animals move in
the direction in which they are facing \cite{bovet}.

For one dimensional systems the probability distribution function for
the position of the walker independent of the direction it is moving,
satisfies the telegrapher's equation in the continuous limit \cite{Weiss}. In two
dimensions, the probability distribution function for the position of
the walker can be shown to satisfy a telegrapher's equation only
approximately, and little is known for three and higher dimensions
\cite{Sev, HL}. However, in all cases, the net effect of persistence is that it
induces a short term memory in the motion in the sense that the
direction of the path is closely related to previous directions over a
correlation time, but in the long run, the transport process remains
diffusive.

An interesting generalization of this process is the case in which the
tendency to move is not in the same direction as the previous step,
but rather, at a fixed deviation from the previous direction. In two
dimensions, this modification induces a tendency to ``loop'' if, say,
the direction of each step is narrowly centered at a fixed, non zero
angle from the previous step. This modification was shown to give rise
to nonmonotonic behavior of the walker's effective diffusion constant
as a function of the width of the distribution of angle changes \cite{HL}. For
very small widths, the walker would perform many loops that brought it
back close to it's initial position and, thus, giving rise to small
diffusion. If, on the other hand, the distribution is very broad, then
the persistence effects are quickly lost and the process is
essentially a normal random walk. However, between both limits, large
loop segments can become uncorrelated before returning near
the starting point, giving rise to a large effective diffusion
coefficient \cite{HL}.

In this work we address the three dimensional case, namely we study
the transport properties of persistent diffusion with curvature and
torsion. This three dimensional motion induces a tendency for the
trajectories to spiral, a behavior that has been observed in many
animals ``\ldots helical trajectories are nearly universal for
organisms that are less than 0.5 mm long and live in water.'' \cite{CrenI}

Further, in spite of its limitations, persistent random walks are
useful models for the shape of linear polymers with some degree of
bond rigidity; thus persistent diffusion with curvature and torsion
would be appropriate models for polymers that tend to form helices
\cite{Flory, Becker}. 

As in \cite{Julicher}, we formulate the problem in terms of the
Frenet--Serret equations with noise terms added to the curvature and
torsion. We then construct a Fokker--Planck equation \cite{Risken} for
this process, from which we can calculate the correlation functions of
the Frenet frame. In particular, we can evaluate the velocity
autocorrelation function from which the effective diffusion constant
of the process is obtained.

\section{Formalism}
\label{first}

To formulate the persistent diffusive process, we consider the Frenet
description of a trajectory \cite{CrenI, CrenII, CrenIII,Julicher},
which consists in the evolution of the right handed frame of unit
vectors given by ${\hat T}$, the vector tangent to the curve, the
normal ${\hat N}$, in the direction of the derivative of ${\hat T}$,
and the binormal ${\hat B}= {\hat T} \times{\hat N}$. The
Frenet--Serret equations then state that
\begin{equation}
\frac{d}{ds}
\left(
  \begin{array}{ccc}
  {\hat T}\\
  {\hat N}\\
  {\hat B}\\
 \end{array}
\right)
=
\left(
\begin{array}{ccc}
  0 & K & 0 \\
  -K & 0 & T \\
  0 & -T & 0 \\
 \end{array}
\right)
\left(
  \begin{array}{ccc}
  {\hat T}\\
  {\hat N}\\
  {\hat B}\\
 \end{array}
\right)
\label{Frenet}
\end{equation}
where $s$ is the arclength, $K$ and $T$ are the curvature and torsion
of the curve respectively, which can, in principle, depend on $s$
too. The time evolution of the particle following the trajectory is
given by
\begin{equation}
\frac{d \vec X}{dt}= v{\hat T}
\end{equation}
where $v$ is the speed of the particle, which we will take as a
constant throughout this work..

To simplify the construction of the Fokker--Planck equation, we begin
by discretizing the Frenet--Serret equation as follows. We write the arclength
as $s=n\Delta s$ and take the following discretized version of the
evolution equation:
\begin{equation}
  \left(
  \begin{array}{ccc}
  {\hat T}\\
  {\hat N}\\
  {\hat B}\\
 \end{array}_{n+1}
\right)
=
R_n
\left(
  \begin{array}{ccc}
  {\hat T}\\
  {\hat N}\\
  {\hat B}\\
 \end{array}_n
\right)
\label{DiscFrenet}\\
\end{equation}
where
\numparts
\begin{eqnarray}
R_n&=&\frac{1+\frac{1}{2}\Omega_n}{1-\frac{1}{2}\Omega_n}\\
\Omega_n &=&
\left(
\begin{array}{ccc}
  0 & K_n & 0 \\
  -K_n & 0 & T_n \\
  0 & -T_n & 0 \\
 \end{array}.
\right)
\end{eqnarray}
\endnumparts
The rationale for this choice is that at each step, the Frenet frame
undergoes a pure rotation. Equivalently, inverting
eq.(\ref{DiscFrenet}) we have
\begin{equation}
  \left(
  \begin{array}{ccc}
  {\hat T}\\
  {\hat N}\\
  {\hat B}\\
 \end{array}_{n}
\right)
=
R_n^{-1}
\left(
  \begin{array}{ccc}
  {\hat T}\\
  {\hat N}\\
  {\hat B}\\
 \end{array}
 \right)_{n+1}
\label{InvDiscFrenet}
\end{equation}
Finally, we introduce noise by taking
\begin{equation}
K_n \to k\Delta s +\sigma_1\eta_1(n)(\Delta s)^{1/2} \qquad
T_n \to \tau\Delta s +\sigma_2\eta_2(n)(\Delta s)^{1/2}
\end{equation}
Where $k$ and $\tau$ will be assumed to be constant, while $\eta_1(n)$ and
$\eta_2(n)$ are random variables satisfying
\begin{equation}
\langle\eta_j(n)\rangle=0 \qquad \langle\eta_j(n)\eta_i(m)\rangle =
\delta_{i,j} \delta_{n,m}
\end{equation}
To order $\Delta s$ we can evaluate the matrix $R_n$ as follows
\begin{eqnarray}
\fl R_n^{-1} \approx 1-\Omega_n+\frac12\Omega_n^2\nonumber\\
\fl\qquad= \left(
\begin{array}{ccc}
1-\frac12 K_n^2& -K_n&K_nT_n\\
K_n&1-\frac12 (K_n^2+T_n^2),& -T_n\\
K_nT_n& T_n&1-\frac12 T_n^2
\end{array}
\right)\nonumber \\
\fl\qquad\approx
\left(
\begin{array}{ccc}
1-\frac{\Delta s}2 \sigma_1^2\eta_1^2& -k\Delta s -\sigma_1\eta_1(\Delta s)^{1/2} &\Delta s\,\sigma_1\sigma_2\eta_1\eta_2\\
k\Delta s +\sigma_1\eta_1(\Delta s)^{1/2} &1-\frac{\Delta s}2 (\sigma_1^2\eta_1^2+\sigma_2^2\eta_2^2)& 
-\tau\Delta s -\sigma_2\eta_2(\Delta s)^{1/2}\\
\Delta s\,\sigma_1\sigma_2\eta_1\eta_2& \tau\Delta s +\sigma_2\eta_2(\Delta s)^{1/2}&1-\frac{\Delta s}2 \sigma_2^2\eta_2^2
\end{array}
\right)
\label{eq:rnapprox}
\end{eqnarray}
where the $n$-dependencies have been suppressed from the last
expression in the interest of typographical clarity. 

Thus, if we denote $P_n({\hat T}, {\hat N}, {\hat B})$ the probability
density function for the Frenet frame being given by the vectors
${\hat T}$, ${\hat N}$ and ${\hat B}$ at step $n$, then, correct to
order $\Delta s$, we derive the Fokker--Planck equation from the relation
\begin{equation}
P_{n+1}({\hat T}, {\hat N}, {\hat B})=
\left\langle P_{n}
\left(
R_n^{-1}({\hat T}, {\hat N}, {\hat B})
\right)
\right\rangle.
\label{eq:master}
\end{equation}
Here the angular brackets denote the average over the random noise $\eta_1(n)$ and $\eta_2(n)$
and $R_n^{-1}$ is as in (\ref{eq:rnapprox}).

%\begin{eqnarray}
%&&P_{n+1}({\hat T}, {\hat N}, {\hat B})= \\
%&&\left\langle P_{n} ( [1-\sigma_1^2\eta_1^2\Delta s/2]{\hat T}
%-[k\Delta s +\sigma_1\eta_1(\Delta s)^{1/2}] {\hat N} +
%\sigma_1\sigma_2\eta_1\eta_2\Delta s {\hat B},\right.\nonumber\\
%&&\qquad
%[k\Delta s +\sigma_1\eta_1(\Delta s)^{1/2}]{\hat T} +
%[1-\sigma_1^2\eta_1^2\Delta s/2 -\sigma_2^2\eta_2^2\Delta s/2]{\hat N}
%-[\tau\Delta s +\sigma_2\eta_2(\Delta s)^{1/2}] {\hat B}, \nonumber\\
%&&\qquad\qquad
%\left. \sigma_1\sigma_2\eta_1\eta_2\Delta s{\hat T}+[\tau\Delta s 
%+\sigma_2\eta_2(\Delta s)^{1/2}] {\hat N}+
%[1-\sigma_2^2\eta_2^2\Delta s/2] {\hat B})\right\rangle \nonumber
%\end{eqnarray}
To lighten the expressions to come, let us now define the drifts
\numparts
\begin{eqnarray}
\vec v_T&=&\frac{\sigma_1^2}{2}{\hat T} + k{\hat N},
\label{eq:drifta}\\
\vec v_N&=&\frac{\sigma_1^2+\sigma_2^2}{2}{\hat N} - k{\hat T} + \tau{\hat
  B},
\label{eq:driftb}
\\
\vec v_B&=&\frac{\sigma_2^2}{2}{\hat B} -\tau {\hat N}.
\label{eq:driftc}
\end{eqnarray}
\label{eq:drifts}
\endnumparts
Putting  in (\ref{eq:master}), $P_n({\hat T}, {\hat N}, {\hat B})= P({\hat T}, {\hat N},
{\hat B}; s)$, with $s=n\Delta s$, expanding to linear order in
$\Delta s$ and finally dividing by $\Delta s$, we obtain
\begin{eqnarray}
\label{fp}
\fl\frac{\partial P}{\partial s}=
-\vec v_T\cdot\nabla_T P
-\vec v_N\cdot\nabla_N P
-\vec v_N\cdot\nabla_B P
\nonumber\\
\quad +\frac{1}{2}\sum\limits_i\sum\limits_j \left[ N_iN_j\left(\sigma_1^2 
\frac{\partial^2 P}{\partial T_i\partial T_j} + \sigma_2^2 
\frac{\partial^2 P}{\partial B_i\partial B_j}\right) +
\left( \sigma_1^2T_iT_j+\sigma_2^2B_iB_j\right)
\frac{\partial^2 P}{\partial N_i\partial N_j}\right]\nonumber\\
\quad-\sum\limits_i\sum\limits_j \left[\sigma_1^2N_iT_j
\frac{\partial^2 P}{\partial T_i\partial N_j} + \sigma_2^2N_iB_j
\frac{\partial^2 P}{\partial B_i\partial N_j}\right]
\end{eqnarray}
%
%\begin{eqnarray}\label{fp}
%\frac{\partial P}{\partial s} = && \\
%&&-\left(\frac{\sigma_1^2}{2}{\hat T} + k{\hat N}\right)\cdot\nabla_T P
%-\left(\frac{\sigma_1^2+\sigma_2^2}{2}{\hat N} - k{\hat T} + \tau{\hat
%  B}\right)\cdot\nabla_N P
%-\left(\frac{\sigma_2^2}{2}{\hat B} -\tau {\hat N}\right)\cdot\nabla_B P
%\nonumber\\
%&&\qquad +\frac{1}{2}\sum\limits_i\sum\limits_j \left[ N_iN_j\left(\sigma_1^2 
%\frac{\partial^2 P}{\partial T_i\partial T_j} + \sigma_2^2 
%\frac{\partial^2 P}{\partial B_i\partial B_j}\right) +
%\left( \sigma_1^2T_iT_j+\sigma_2^2B_iB_j\right)
%\frac{\partial^2 P}{\partial N_i\partial N_j}\right]\nonumber\\
%&&\qquad \qquad-\sum\limits_i\sum\limits_j \left[\sigma_1^2N_iT_j
%\frac{\partial^2 P}{\partial T_i\partial N_j} + \sigma_2^2N_iB_j
%\frac{\partial^2 P}{\partial B_i\partial N_j}\right] \nonumber
%\end{eqnarray}
where $\nabla_T$, $\nabla_N$ and $\nabla_B$ denote the gradient in the
$T$, $N$ and $B$ coordinates respectively. The equation above
describes the evolution of the probability distribution for the
components of the Frenet frame along the arclength of the particle's
trajectory and becomes exact in the limit $\Delta s \to 0$. Equation
(\ref{fp}) can be rewritten in a slightly more symmetric form as
\begin{eqnarray}
\frac{\partial P}{\partial s} = 
-\vec v_T\cdot\nabla_T P
-\vec v_N\cdot\nabla_N P
-\vec v_N\cdot\nabla_B P\nonumber\\
\qquad +\frac{\sigma_1^2}{2}\sum\limits_i\sum\limits_j \left( 
N_i\frac{\partial}{\partial T_i}
-T_i\frac{\partial}{\partial N_i}\right) \left( 
N_j\frac{\partial}{\partial T_j}
-T_j\frac{\partial}{\partial N_j}\right) P \nonumber \\
\qquad +\frac{\sigma_2^2}{2}\sum\limits_i\sum\limits_j \left( 
N_i\frac{\partial}{\partial B_i}
-B_i\frac{\partial}{\partial N_i}\right) \left( 
N_j\frac{\partial}{\partial B_j}
-B_j\frac{\partial}{\partial N_j}\right)P;
\end{eqnarray}
%\begin{eqnarray}
%\frac{\partial P}{\partial s} = &&\\
%&&-\left(\frac{\sigma_1^2}{2}{\hat T} + k{\hat N}\right)\cdot\nabla_T P
%-\left(\frac{\sigma_1^2+\sigma_2^2}{2}{\hat N} - k{\hat T} + \tau{\hat
%  B}\right)\cdot\nabla_N P
%-\left(\frac{\sigma_2^2}{2}{\hat B} -\tau {\hat N}\right)\cdot\nabla_B P
%\nonumber\\
%&&\qquad +\frac{\sigma_1^2}{2}\sum\limits_i\sum\limits_j \left( 
%N_i\frac{\partial}{\partial T_i}
%-T_i\frac{\partial}{\partial N_i}\right) \left( 
%N_j\frac{\partial}{\partial T_j}
%-T_j\frac{\partial}{\partial N_j}\right) P \nonumber \\
%&&\qquad +\frac{\sigma_2^2}{2}\sum\limits_i\sum\limits_j \left( 
%N_i\frac{\partial}{\partial B_i}
%-B_i\frac{\partial}{\partial N_i}\right) \left( 
%N_j\frac{\partial}{\partial B_j}
%-B_j\frac{\partial}{\partial N_j}\right); P \nonumber
%\end{eqnarray}
either way, the equation seems to be rather intractable. However, it
can be used to study the evolution of the moments of ${\hat T}$,
${\hat N}$ and ${\hat B}$; from these, the transport properties of $\vec
x(t)$ can be inferred. In particular, by carrying out the integrations,
one finds that
\begin{equation}
\frac{d}{ds}
  \left(
  \begin{array}{ccc}
 \langle {\hat T}_s|{\hat T_0,\hat  N_0, \hat B_0} \rangle \\
 \langle {\hat N}_s|{\hat T_0, \hat N_0, \hat B_0} \rangle \\
 \langle {\hat B}_s|{\hat T_0,\hat  N_0,\hat  B_0 } \rangle  \\
 \end{array}
\right)
=
\left(
\begin{array}{ccc}
 -\sigma_1^2/2 & k & 0 \\
  -k & -[\sigma_1^2+\sigma_2^2]/2 & \tau \\
  0 & -\tau & -\sigma_2^2/2 \\
 \end{array}
 \begin{array}{ccc}
 \langle {\hat T}_s|{\hat T_0,\hat N_0, \hat B_0 } \rangle \\
 \langle {\hat N}_s|{\hat T_0,\hat N_0, \hat B_0 } \rangle \\
 \langle {\hat B}_s|{\hat T_0,\hat N_0, \hat B_0} \rangle \\
 \end{array}
 \right).
\label{Means}
\end{equation}
Here the dependence on the initial condition has been written
explicitly to emphasize the fact that these statistics give us access to the
time correlation functions. The solution can be written as
\begin{equation}
\left(
  \begin{array}{ccc}
 \langle {\hat T}_s|{\hat T_0,\hat  N_0,\hat  B_0} \rangle \\
 \langle {\hat N}_s|{\hat T_0,\hat  N_0,\hat  B_0} \rangle \\
 \langle {\hat B}_s|{\hat T_0, \hat N_0, \hat B_0 } \rangle  \\
 \end{array}
 \right)
= e^{\Gamma s}
  \left(
  \begin{array}{ccc}
  {\hat T_0}\\
  {\hat N_0}\\
  {\hat B_0}\\
 \end{array}
 \right)
\end{equation}
where $\Gamma$ is the matrix appearing in eq.(\ref{Means}). Thus, the
matrix of correlation functions can be written explicitly as
\begin{equation}
\mathcal{M}(s) =
 \left(
 \begin{array}{ccc}
  \langle \hat T_s\cdot \hat T_0 \rangle &  \langle \hat T_s\cdot \hat
  N_0 \rangle & \langle \hat T_s\cdot \hat B_0 \rangle \\
   \langle \hat N_s\cdot \hat T_0 \rangle &  \langle \hat N_s\cdot \hat
  N_0 \rangle & \langle \hat N_s\cdot \hat B_0 \rangle \\
  \langle \hat B_s\cdot \hat T_0 \rangle &  \langle \hat B_s\cdot \hat
  N_0 \rangle & \langle \hat B_s\cdot \hat B_0 \rangle 
 \end{array}
 \right)
= e^{\Gamma s}
\end{equation}
The Laplace-Fourier transform of the correlation matrix is thus given by
\begin{equation}
\mathcal{M}(\omega)= (i\omega - \Gamma)^{-1}.
\end{equation}
If we now define $\Delta(\omega)$ as follows
\begin{equation}
\Delta(\omega)=(\sigma_1^2+2i\omega) (\sigma_2^2+2i\omega)(\sigma_1^2+ 
\sigma_2^2+2i\omega)+4k^2(\sigma_2^2+2i\omega)+4\tau^2(\sigma_1^2+2i\omega),
\label{eq:delta}
\end{equation}
we find the matrix $\Delta(\omega){\cal M}(\omega)/2$ to be given by
\begin{equation}
\fl\left(
\begin{array}{lll}
 (\sigma_2^2+2i\omega)(\sigma_1^2+\sigma_2^2+2i\omega)+4\tau^2 & 
2k(\sigma_2^2+2i\omega) & 4k\tau \\
-2k(\sigma_2^2+2i\omega) & (\sigma_1^2+2i\omega) (\sigma_2^2+2i\omega) &
2\tau(\sigma_1^2+2i\omega)\\
4 k\tau & -2\tau(\sigma_1^2+2i\omega)& (\sigma_1^2+2i\omega)(\sigma_1^2
 +\sigma_2^2+2i\omega)+4k^2
 \end{array}
 \right)
 .
 \label{3dmatrix}
\end{equation}
Note that the matrix is not symmetric as the process is not invariant
under simple time inversion. Rather, $\hat T$ and $\hat B$ are odd,
whereas $\hat N$ is even, leading to the observed alternation between
symmetric and antisymmetric elements in the matrix.
%\begin{eqnarray}
%&&\mathcal{M}(\omega)= (i\omega - \Gamma)^{-1} \nonumber\\
%&&=\frac{2}{(\sigma_1^2+2i\omega) (\sigma_2^2+2i\omega)(\sigma_1^2+ 
%\sigma_2^2+2i\omega)+4k^2(\sigma_2^2+2i\omega)+4\tau^2(\sigma_1^2+2i\omega)}
%  \times \\
%&&\nonumber\\
%&&
%\qquad
%\begin{array}{ccc}
% (\sigma_2^2+2i\omega)(\sigma_1^2+\sigma_2^2+2i\omega)+4\tau^2 & 
%2k(\sigma_2^2+2i\omega) & 4k\tau \\
%-2k(\sigma_2^2+2i\omega) & (\sigma_1^2+2i\omega) (\sigma_2^2+2i\omega) &
%2\tau(\sigma_1^2+2i\omega)\\
%4 k\tau & -2\tau(\sigma_1^2+2i\omega)& (\sigma_1^2+2i\omega)(\sigma_1^2
% +\sigma_2^2+2i\omega)+4k^2
% \end{array}\label{3dmatrix}\nonumber
%\end{eqnarray}
\section{Special cases and particular results}
The explicit inversion of the above results to obtain the
time-dependent correlation functions can, of course, be carried out in
principle, as it only requires the solution of a third order
polynomial. Nevertheless, the results are messy and
unenlightening. Instead, we focus on particular cases which illustrate
different aspects of the process. One simple situation is reached in
the planar case, in which both the torsion $\tau$ and its associated
noise vanish ($\sigma_2 =0$). In this case, the binormal vector is
constant and the motion is restricted to a plane. In this situation,
(\ref{3dmatrix}) simplifies to
\begin{equation}
\fl\quad\mathcal{M}_{2d}(\omega)
=\frac{2}{\Delta(\omega)}
\left(
\begin{array}{ccc}
 2i\omega(\sigma_1^2+2i\omega) & 4i\omega k & 0 \\
-4i\omega k & 2i\omega(\sigma_1^2+2i\omega) &
0\\
0 & 0 & (\sigma_1^2+2i\omega)^2+4k^2
 \end{array}
 \right)
 ,\label{2dmatrix}\nonumber
\end{equation}
where $\Delta(\omega)$ here reduces to
\begin{equation}
\Delta_{2d}(\omega)=2i\omega[(\sigma_1^2+2i\omega)^2+4k^2].
\end{equation}
%\begin{eqnarray}
%&&\mathcal{M}_{2d}(\omega) \\
%&&\qquad =\frac{1}{i\omega[(\sigma_1^2+2i\omega)^2+4k^2]}
%  \times
%\begin{array}{ccc}
% 2i\omega(\sigma_1^2+2i\omega) & 4i\omega k & 0 \\
%-4i\omega k & 2i\omega(\sigma_1^2+2i\omega) &
%0\\
%0 & 0 & (\sigma_1^2+2i\omega)^2+4k^2
% \end{array}\label{2dmatrix}\nonumber
%\end{eqnarray}
The $s$-dependent correlations can also be determined explicitly in this case. Indeed
one readily finds
\begin{equation}
e^{\Gamma s}=\left(
\begin{array}{ccc}
\cos(ks)e^{-\sigma_1^2s/2}&\sin(ks)e^{-\sigma_1^2s/2}&0\\
-\sin(ks)e^{-\sigma_1^2s/2}&\cos(ks)e^{-\sigma_1^2s/2}&0\\
0&0&1
\end{array}
\right)
\label{eq:dyn2d}
\end{equation}
which corresponds to circular motion which is damped due to the noise.

Further, we can use $\mathcal{M}$ to write the Fourier-Laplace transforms for
the average of the Frenet vectors; for example, for the mean tangent
vector we have:
\begin{eqnarray}
\fl \langle {\hat T}_\omega|{\hat T_0,\hat N_0,\hat B_0} \rangle = \frac2{\Delta(\omega)}
\bigg\{
\left[
(\sigma_2^2+2i\omega)(\sigma_1^2+\sigma_2^2+2i\omega)+4\tau^2
\right]\nonumber\\
\hat T_0 + 2k(\sigma_2^2+2i\omega)\hat N_0 + 4k\tau \hat B_0
\bigg\}.
\end{eqnarray}
This quantity connects to the behavior of the walker's average position by
\begin{eqnarray}
\langle {\vec x}_\omega|{\hat T_0,\hat N_0,\hat B_0}\rangle =\frac{1}{i\omega}\langle {\hat T}_\omega|{\hat T_0,\hat N_0,\hat B_0}\rangle.
\end{eqnarray}
Thus, in the long time limit, the average position of the persistent walker
saturates at
\begin{eqnarray}
\fl\langle {\vec x}_t|{\hat T_0,\hat N_0,\hat B_0}\rangle \to
\langle {\vec x}_\infty|{\hat T_0,\hat N_0,\hat B_0}\rangle =
2 \frac{\left\{\sigma_2^2(\sigma_1^2+\sigma_2^2)+4\tau^2\right\} \hat T_0 
+ 2k\sigma_2^2\hat N_0 + 4k\tau \hat B_0}{\sigma_1^2 \sigma_2^2(\sigma_1^2+
  \sigma_2^2)+4k^2\sigma_2^2+4\tau^2\sigma_1^2}.
\label{Posit}
\end{eqnarray}
Another interesting quantity is the Darboux vector \cite{CrenI}, which in our case,
will be writen as $\vec A =\tau \hat T + k\hat B$. This vector is akin
to the angular momentum of the walker. In the absence
of noise, this vector is constant, as can be seen directly from
(\ref{Means}), and lies on the axis of the helical motion. In the
presence of noise, however, the Darboux vector evolves according to
\begin{eqnarray}
\fl\quad
\langle {\vec A}_\omega|{\hat T_0,\hat N_0,\hat B_0} \rangle &=&
\frac{2}{\Delta(\omega)}\bigg\{(\sigma_1^2+\sigma_2^2+2i\omega)\left(
\sigma_2^2 \tau \hat T_0 + \sigma_1^2k \hat B_0\right) +
\nonumber\\
&&  2\left[
i\omega
\left(
\sigma_1^2+\sigma_2^2+2i\omega
\right)
+2(k^2+\tau^2)\right]\vec A_0 
+2k\tau
\left(
\sigma_2^2-\sigma_1^2
\right)
 \hat N_0
\bigg\}.
\end{eqnarray}
%\begin{eqnarray}
%&& \langle {\vec A}_\omega|{\hat T_0,\hat N_0,\hat B_0} \rangle =\\
%&& 2\frac{(\sigma_1^2+\sigma_2^2+2i\omega)\left[
%(\sigma_2^2 \tau \hat T_0 + \sigma_1^2k \hat B_0\right] + 2\left[
%i\omega(\sigma_1^2+\sigma_2^2+2i\omega)+2(k^2+\tau^2)\right]\vec A_0 
%+2k\tau\left[\sigma_2^2-\sigma_1^2\right] \hat N_0}
%{(\sigma_1^2+2i\omega) (\sigma_2^2+2i\omega)(\sigma_1^2+
%  \sigma_2^2+2i\omega)+4k^2(\sigma_2^2+2i\omega)+4\tau^2(\sigma_1^2+2i\omega)}
%\nonumber
%\end{eqnarray}
Here $\vec A_0=\tau \hat T_0 + k\hat B_0$ is the initial value of
$\vec A$. It is worth noting that in the small noise limit, the
saturation value of the average position, given in (\ref{Posit}) is given by
\begin{eqnarray}
\langle {\vec x}_\infty|{\hat T_0,\hat N_0,\hat B_0}\rangle
\approx \frac{2\tau \vec A_0}{ \sigma_2^2 k^2 +\sigma_1^2\tau^2}
\end{eqnarray}
which reflects the intuitively appealing fact that in the low-noise limit, the net
displacement is in the direction of the axis of the helix.

% The other relatively simple situation arises when $\sigma_1=\sigma_2$.

Perhaps more interestingly, from the correlation functions we can
compute the effective diffusion constant. Assuming, again, that the
instantaneous speed  $v$ is constant, we can write
\begin{equation}
\left\langle
 \vec x(t)\cdot\vec x(t)
\right\rangle = 
v^2 \int\limits_0^t\int\limits_0^t 
\left\langle
 \hat T_{s_1}\cdot
\hat T_{s_2}
\right\rangle\,dt_1\,dt_2,
\end{equation}
recalling that $s_i=vt_i$. The effective diffusion constant $D$ will be given by
\begin{equation}
\fl 
D=\lim\limits_{t\to\infty} \frac{
\left\langle
 \vec x(t)\cdot\vec
  x(t) 
  \right\rangle}
  {2t}=v\int\limits_0^\infty 
  \left\langle
  \hat T_{s}\cdot
\hat T_0
\right\rangle
\,ds = 2v\,\frac{\sigma_2^2(\sigma_1^2 + \sigma_2^2) +
4\tau^2}
{ \sigma_1^2\sigma_2^2 (\sigma_1^2 + \sigma_2^2) +
4k^2\sigma_2^2 + 4\tau^2\sigma_1^2}
\label{eq:diff}
\end{equation}
This expression's behaviour becomes easier to visualize after introducing scaled
versions of the curvature, the torsion and the diffusion constant as well as the 
adimensional parameter $\kappa$, which measures the relative importance of the curvature and the torsion:
\numparts
\begin{eqnarray}
\overline{k}&=&\frac{k}{\sigma_1\sqrt{\sigma_1^2+\sigma_2^2}}
\label{eq:scaleda}\\
\overline{\tau}&=&\frac{\tau}{\sigma_2\sqrt{\sigma_1^2+\sigma_2^2}}
\label{eq:scaledb}\\
\overline{D}&=&\frac{\sigma_1^2}{2v}\,D
\label{eq:scaledc}\\
\kappa&=&\frac{4\overline{k}^2}{1+4\overline{\tau}^2}
\label{eq:scaledd}
\end{eqnarray}
\label{eq:scaled}
\endnumparts
In terms of these parameters, (\ref{eq:diff}) then reduces to 
\begin{equation}
\overline{D}=\frac{1+4\overline{\tau}^2}{1+4\overline{k}^2+4\overline{\tau}^2}=\frac1{1+\kappa}
\label{eq:diff1}
\end{equation}
Thus $\overline{D}$ grows with $\overline{\tau}$ and decreases with
$\overline{k}$.  Intuitively, this corresponds to the fact that
torsion favors a rectilinear motion in the direction of the Darboux
vector, thus favouring diffusion by enhancing transport, whereas
curvature leads to looping around in circles, which confines the
walker's motion. For the true diffusion constant, matters are a bit
more complex, since it depends really on three parameters.  Thus, for
instance, upon increasing $\sigma_1$ at fixed
$\sigma_1^2+\sigma_2^2$---that is, at fixed total noise---we decrease
$\overline{k}$ and thus increase $\overline{D}$. The behaviour of $D$,
however, is in general non-monotonic, since it corresponds to
$\overline{D}$ {\em divided\/} by $\sigma_1^2$.

It is interesting to note that(\ref{eq:diff}) recovers the two-dimensional result
taking the torsion $\tau\to0$ first, simplifying and then $\sigma_2\to
0$. This procedure yields
\begin{equation}
D_{2d}= \frac{2\sigma_1^2}{\sigma_1^4+ 4k^2}\,v
\label{eq:diff2d}
\end{equation}
which is analogous to the result obtained in \cite{HL}. We note that,
as above, $D_{2d}$ is a non monotonic function of $\sigma_1$. It is
also interesting to note that a different, intrinsically three
dimensional result, is obtained if we take the limits in opposite
order. Indeed, if we let $\sigma_2\to 0$ first, we see that
$\kappa\to0$ and therefore
\begin{equation}
D_{H}= \frac{2}{\sigma_1^2}\,v
\label{eq:diffweird}
\end{equation}
which is independent of all the other parameters of the
problem. From the scaling form (\ref{eq:diff1}), it follows quite 
straightforwardly that the same result also holds if $k=0$. Intuitively, 
however, it is quite far from being obvious why, in  the former case, the
diffusion constant only depends on $\sigma_1$ and $v$. 

Indeed, it is quite clear that a non-trivial crossover behaviour must
be observed when we have $\sigma_2=0$ and $\tau$ small. In this case,
we might have expected the behavior to be close to that for $\tau=0$,
yet the exact expressions for the diffusion constants differ by a
finite amount. This means that, in this case, we shall observe
two-dimensional diffusion for a long-time, and would expect to see the
corresponding diffusion constant, see (\ref{eq:diff2d}) over a limited
time range. After that, we would expect torsion to bring the system
significantly out of the plane, and thus to lead to the expression
(\ref{eq:diffweird}). Indeed, this can be verified by a
straightforward though tedious analysis of the zeroes of
$\Delta(\omega)$. One finds a low-lying imaginary zero with the value
\begin{equation}
\omega_0=\frac{2i\sigma_1^2\tau^2}
{\sigma_1^4+4k^2}
\end{equation}
which corroborates the picture

When $k=0$, on the other hand, the tendency of the process is to move,
on average, in the same direction of the previous motion, so that we
have a typical persistent walk. The torsion, on the other hand, with
or without noise, merely rotates the new step around the previous
tangent and thus has no influence on the diffusive properties of the
process.

\section{Asymptotic shape of the average trajectory in the low-noise limit}

As a final remark, we observe a peculiar transition in the large-time
behaviour of the average trajectory. Indeed, the average position of
the trajectory at arclength $s$ is determined by the matrix
$\exp(\Gamma s)$. For large $s$ it must therefore be dominated by that
eigenvalue of $\Gamma$ having the smalles real part in absolute value.
As is readily seen, the three eigenvalues of $\Gamma$ in the limit of
small noise are given by
\numparts
\begin{eqnarray}
\lambda_0&=&-\frac{\tau^2\sigma_1^2+k^2\sigma_2^2}{2(k^2+\tau^2)}
\label{eq:eigena}\\
\lambda_{\pm}&=&-\frac{\sigma_1^2(2k^2+\tau^2)+\sigma_2^2(k^2+2\tau^2)}{4(k^2+\tau^2)}\pm i\sqrt{k^2+\tau^2}
\label{eq:eigenb}
\end{eqnarray}
\label{eq:eigen}
\endnumparts
up to terms of higher order in the noise. 
Thus, in the low-noise limit, we have two groups of eigenvalues, one
tending to $0$ and corresponding essentially to 
the Darboux vector as an eigenvector. This thus describes the motion in the direction of the helix axis. 
The other two eigenvalues correspond to eigenvectors determining the rotational motion. 
As we see, their real parts are not equal, and an easy calculation shows that $\lambda_0$ dominates if
\begin{equation}
\Phi=(2k^2-\tau^2)\sigma_1^2+(2\tau^2-k^2)\sigma_2^2>0
\label{eq:border}
\end{equation}
and $\lambda_\pm$ otherwise. Thus, according to the sign of $\Phi$, we
have two qualitatively different motions: if $\Phi>0$, the circular
motion decays more rapidly than the progressive motion, and the
asymptotic shape of the particle's trajectory is a very loosely wound
helical motion, in which the particle in one rotation advances by a
distance far larger than the helix' radius. If $\Phi<0$, on the other
hand, the opposite situation prevails, and the motion is on a circle
of a radius much larger than the distance by which the particle
advances for each rotation. In other words, we might say that the
asymptotic pitch of the average motion diverges at large times for
$\Phi>0$, whereas it vanishes for $\Phi<0$. We show in Figure 1 the
autocorrelation function of the tangent vector $\hat T$ for large
times for 2 cases in which $\Phi>0$ and $\Phi<0$ respectively. The
difference is quite striking.

Finally, let us point out that the above-mentioned transition does not
always occur. In particular, if $k$ and $\tau$ are equal, or of
similar size, then we always have $\Phi>0$. For both signs of $\Phi$
to be possible, we require
\begin{equation}
\max\left(
\frac{k}{\tau},\frac{\tau}{k} 
\right)\geq\sqrt2.
\label{eq:max}
\end{equation}
When $\Phi=0$, the model maintains a truly helicoidal dynamics for all times. It would be of interest 
to know whether any particular significance attaches to this fact in real systems .
\begin{center}
\includegraphics{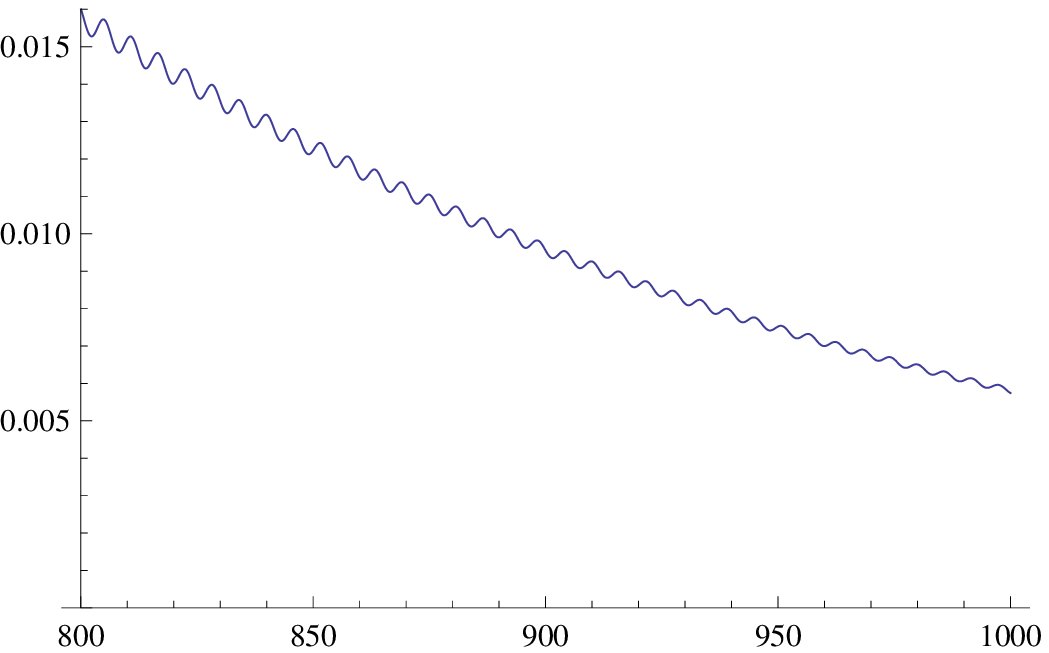}
\includegraphics{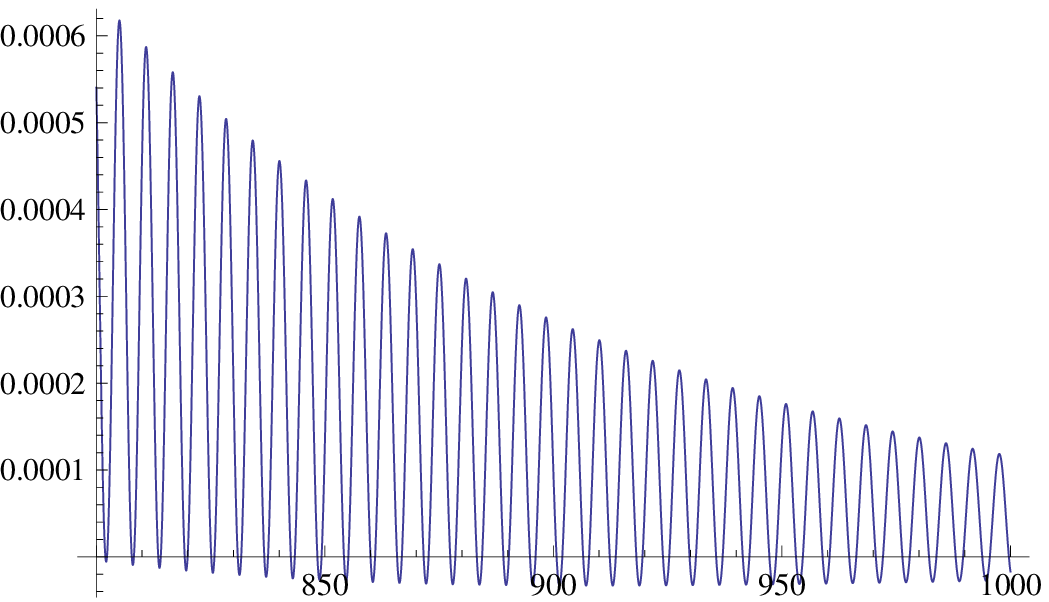}
\end{center}
Fig.~1: The autocorrelatin function for $\hat T$ for $k=0.4$ and $\tau=1$ in the cases where $\sigma_1=\sigma_2=0.1$
(upper figure) and $\sigma_1=0.15$, $\sigma_2=0.05$ (lower figure). 
\section{Conclusions}

Let us briefly summarize what we have done: we have defined a
discretization of the Frenet--Serret equations describing a curve in
three-dimensional space. Assuming that the curvature and torsion are
both constant, and proportional to the discretization parameter
$\Delta s$ this equation we have defines a discretized helical
motion. We add uncorrelated Gaussian noise of the order of $(\Delta
s)^{1/2}$ to the curvature and torsion to define a helical brownian
motion. Next derive the Fokker--Planck equation for this stochastic
process. The resulting process is, in fact, a random walk
in which the probability of every step depends on the step's relative
position with respect to the preceding {\em two\/} steps. Indeed, the
curvature dictates the average angle to the preceding step, whereas
the torsion determines the average angle with respect to the plane
generated by the two earlier steps. The process is fully determined by
five parameters: the average curvature $k$, the average torsion
$\tau$, the noise intensity for the curvature $\sigma_1^2$, the noise
intensity for the torsion $\sigma_2^2$ and finally the speed $v$ at
which the particle moves. While this last parameter only fixes an
otherwise arbitrary time scale and can thus in principle be
disregarded, the other parameters are all essential.

We then proceeded to determine the Fourier--Laplace transform of the
various correlation functions of the orthogonal triad $\hat T$, $\hat
N$ and $\hat B$. From this it is possible to determine the effective
diffusion constant. An explicit formula for it is derived in terms of
the model's five parameters, see (\ref{eq:diff}).  Interestingly,
there is a scaling parameter $\kappa$ defined in terms of these five,
see (\ref{eq:scaled}), such that an appropriately rescaled diffusion
constant depends only on $\kappa$, see (\ref{eq:diff1}).

Other quantities which can be evaluated explicitly are the asymptotic
average position (as a vector) of the walk, as well as the correlation
function of the Darboux vector, which corresponds to the axis of the
helix which would be generated in the noiseless limit. We show in
particular that the asymptotic average position of the walk indeed
coincides with the Darboux vector in the low-noise limit.

Various special cases can also be treated more or less
explicitly. Among these are the two-dimensional case ($\sigma_2=0$ and
$\tau=0$), for which we have given an explicit expression for the
correlations.

A still open question concerns the response of the process when an
external field is applied; the difficulties for advancing in this
question arise from the fact that a spatially fixed field rotates in
the Frenet frame. A related problem would be to understand the effects
in the transport properties due to periodic or strategic variations
in the parameters, attempting to steer the particle in a given
direction \cite{CrenI, CrenII, CrenIII, Julicher}.

\section{Acknowledgements}
FL acknowledges the support of UNAM-PAPIIT-DGAPA 
grant IN114014 as well as CONACyT grant 154586.


\section*{References}

\begin{thebibliography}{99}

\bibitem{CrenI} Crenshaw H C ``Orientation by helical motion--I.  Kinematics of the
  helical motion of organisms with up to six degrees of freedom'' Bulletin of 
  Mathematical Biology {\bf55}  (1) 197--212 1993

\bibitem{CrenII}Crenshaw H C and  Edelstein-Keshet L  ``Orientation by helical motion--II.  Changing the direction
  of the axis of motion''
  Bulletin of Mathematical Biology  {\bf55}  (1) 213--230 1993

\bibitem{CrenIII}Crenshaw H C  ``Orientation by helical motion--III.  Microorganisms can
  orient to stimuli by changing the direction of their rotational
  velocity'' Bulletin of Mathematical Biology
Bulletin of Mathematical Biology  {\bf55}  (1) 231--255 1993  

\bibitem{Flory}  Flory, P J  {\em Statistical Mechanics of Chain Molecules}
Interscience, New York, 1969

\bibitem{Julicher} Friedrich B M  and J\"ulicher F  ``Steering Chiral Swimmers along Noisy Helical Paths
Phys. Rev. Lett. {\bf103} 068102 2009


\bibitem{Becker} Becker N B and Everaers R ``From rigid base pairs to semiflexible polymers:
  Coarse-graining DNA''  Phys. Rev. E {\bf76} 021923 2007

\bibitem{Weiss} Weiss G H  {\em Aspects and Applications of the Random Walk}
  (North-Holland, Amsterdam, 1994, and references therein.

\bibitem{bovet} Bovet P and Benhamou, S  ``Spatial analysis of animals' movements 
using a correlated random walk model'' Journal of theoretical biology
{\bf131} (4), 419--433 1988

\bibitem{Sev}Sevilla F J and G\'omez Nava A  ``Theory of diffusion of active particles that move at
  constant speed in two dimensions''  Phys. Rev. E {\bf90} 022130 2014

\bibitem{HL} Larralde H ``Transport properties of a two-dimensional  `chiral'
  persistent random walk'' Phys. Rev. E {\bf56} 5004 (1997)
\bibitem{Risken} Risken H {\em The Fokker--Planck Equation} Springer-Verlag,
  Berlin 1989




\end{thebibliography}
\end{document}